\documentclass{jkas}

\def\year{2015} 
\def\month{March} 
\def\volume{00} 
\def\issue{0} 
\def\beginpage{1} 
\def\endpage{99} 
\def\received{---} 
\def\accepted{---} 

\date{Received \received ; accepted \accepted}

\title{
Rigorous ``Rich Argument'' in Microlensing Parallax
}

\author[1,2]{Andrew~Gould}

\affil[1]{Max-Planck-Institute for Astronomy, K\"{o}nigstuhl 17, 69117 Heidelberg, Germany}
\affil[2]{Department of Astronomy Ohio State University,
140 W.\ 18th Ave., Columbus, OH 43210, USA 
\email{gould@astronomy.ohio-state.edu }}

\def\corrauthor{
Andrew Gould
}

\def\runningauthor{
Andrew Gould
}

\def\runningtitle{
``Rich Argument''
}

\def\keywords{
gravitational microlensing 
}

\def\abstracttext{
I show that when the observables $(\bpi_\e,t_\e,\theta_\e,\pi_s,\bmu_s)$
are well measured up to a discrete degeneracy in the microlensing
parallax vector $\bpi_\e$, the relative likelihood of the
different solutions can be written in closed form $P_i = K H_i B_i$,
where $H_i$ is the number of stars (potential lenses) having
the mass and kinematics of the inferred parameters of solution $i$
and $B_i$ is an additional factor that is formally derived from
the Jacobian of the transformation from Galactic to microlensing
parameters.  The Jacobian term $B_i$ constitutes an explicit
evaluation of the ``Rich Argument'', i.e., that there is an extra
geometric factor disfavoring large-parallax solutions in addition
to the reduced frequency of lenses given by $H_i$.  Here $t_\e$ is the Einstein
timescale, $\theta_\e$ is the angular Einstein radius, and 
$(\pi_s,\bmu_s)$ are the (parallax, proper motion) of the microlensed
source.  I also
discuss how this analytic expression degrades in the presence of
finite errors in the measured observables.
}

\newcommand{\bdv}[1]{\mbox{\boldmath$#1$}}

\def\au{{\rm AU}}

\def\mas{{\rm mas}}

\def\rel{{\rm rel}}

\def\lim{{\rm lim}}
\def\sat{{\rm sat}}
\def\e{{\rm E}}
\def\bpi{{\bdv\pi}}
\def\bmu{{\bdv\mu}}

\def\bTheta{{\bdv\theta}}

\def\apj{{ApJ}}
\def\aj{{AJ}}
\def\apjl{{ApJL}}

\def\aap{{A\&A}}

\def\mnras{{MNRAS}}

\begin{document}
\jkashead 


\section{{Introduction}
\label{sec:intro}}

The ``Rich argument'' played an important role in motivating 
space-based microlensing studies.  When \citet{refsdal66} introduced
space-based microlensing parallax, he already realized that it would
yield four degenerate solutions in what we now call the microlensing
parallax vector,
\begin{equation}
\bpi_\e = {\pi_\e\over \mu}\bmu,
\qquad \pi_\e = {\pi_\rel\over\theta_\e},
\label{eqn:piedef}
\end{equation}
where $\theta_\e$ is the Einstein radius and $(\pi_\rel,\bmu)$ are the 
lens-source relative (parallax, proper motion).  See Figure~1 of 
\citet{gould94} for an illustration of how this degeneracy arises
and Figure~1 of \citet{ob140939} for the first practical example.

This problem initially appeared as quite severe: in the great majority
of cases for which the actual value of $\pi_\e$ was small (e.g.,
$\pi_\e \lesssim 0.1$), there would be an alternate solution in which it
was large (e.g., $\pi_\e\sim 1$).  That is, the microlens parallax is
given by
\begin{equation}
\bpi_\e = {\au\over D_\perp}
\biggl({t_{0,\sat} - t_{0,\oplus}\over t_\e},u_{0,\sat} - u_{0,\oplus}\biggr),
\label{eqn:dtaudbeta}
\end{equation}
where $(t_0,u_0)$ are the time of peak and impact parameter
as seen from either Earth
or the satellite, $t_\e=\theta_\e/\mu$ is the Einstein timescale, and
${\bf D}_\perp$ is the projected satellite-Earth separation vector.
Because $u_0$ is a signed quantity for which only the absolute value is normally
measured, events whose true $u_0$ values are 
$(u_{0,\sat},u_{0,\oplus})=(0.5,0.4)$ can also be interpreted as 
$(u_{0,\sat},u_{0,\oplus})=(0.5,-0.4)$  and hence (according to 
Equation~(\ref{eqn:dtaudbeta})) with a second parallax component that
is nine times larger.
Then, because $\pi_\e$ enters directly
into the mass and distance estimates,
\begin{equation}
M = {\theta_\e\over\kappa\pi_\e};
\quad
\pi_\rel = \theta_\e\pi_\e;
\quad
\kappa = {4G\over c^2\au}\simeq 8.14{\mas\over M_\odot},
\label{eqn:massdis}
\end{equation}
this degeneracy appeared to pose a major obstacle to the interpretation
of any space-based microlensing experiment.

\citet{refsdal66} had already proposed a ``simple'' solution: launch
a second satellite into solar orbit to take simultaneous observations.
See also Figure~4 of \citet{gould19}.  However, given the challenges
(in the first place, the expense) of launching even one such satellite,
this did not appear as a practical approach.

James Rich (circa 1997, private communication) argued that these
alternate solutions were geometrically improbable.  Hence, while they
could not be ruled out in any particular case, their presence would not
interfere with the statistical interpretation of a parallax-satellite
experiment.  This insight had an important motivating impact on early
workers who were investigating the mathematical and physical basis of
microlensing parallax.

However, this argument only became widely known when, following the first
large-scale satellite-parallax campaign \citep{ob140124,ob140939} using 
the {\it Spitzer} satellite in solar orbit, \citet{21event} 
explicitly gave this argument and made the first attempt to quantify it
in the course of analyzing 21 {\it Spitzer} events from 2014.  Based on purely
geometric reasoning, they argued that larger-parallax solutions were
disfavored by $\pi_\e^{-2}$.

It was always known that, in the absence of any other information,
large parallax (i.e., nearby-lens) solutions were disfavored simply
because of the smaller volume available.  And also that this effect
was often augmented by the lower space density of stars for the 
more nearby solution.  However, the ``Rich argument'' was regarded
(correctly, as we will see) as additionally disfavoring the 
large-parallax solutions.

In fact, \citet{mb09387} already explicitly noted such an effect in their
analysis of MOA-2009-BLG-387, for which purely ground-based data
yielded a measurement of $\bpi_\e$ with large error bars.  When they
estimated $\bpi_\e$ by 
weighting the microlensing likelihood of each $\bpi_\e$ value 
according to a prior based on a Galactic model, they found additional
purely geometric terms in the Jacobian arising from the transformation 
from physical coordinates to microlensing parameters.  See their
Equations~(17)--(18).  They then showed (lower-left panel of their
Figure~6) that the combination of Galactic and Jacobian factors
drove the solution 2--3 $\sigma$ from the best fit based only on
the $\chi^2$ of the microlensing fit.

Mathematically, the ``discrete degeneracies'' (multiple isolated maxima
in the likelihood function) that appear routinely in space-based
microlensing parallax are simply a special case of a more general likelihood
function, such as the one analyzed by \citet{mb09387}.  Therefore, 
exactly the same Galactic factors and Jacobian factors should appear
in both.  However, while this statement would probably have appeared 
obvious if it had been so formulated, it was not made initially.  
Hence, for example,
\citet{zhu17} considered two prescriptions for weighting discrete
$\bpi_\e$ solutions in their analysis of 50 {\it Spitzer} events from
2015.  In one, they calculated the likelihood of solutions using a 
product of the light-curve likelihood and Galactic-model likelihood.
In the second, they further multiplied by $\pi_\e^{-2}$ for the ``Rich argument''.
Their Galactic model was implemented by numerical integration over
physical parameters, and hence it implicitly contained the Jacobian
factors discussed above.  However, as a practical matter they found
that their statistical results depended only weakly on this choice.
In their study of the \citet{zhu17} results, \citet{kb19} argued 
that the ``Rich argument'' was simply an ad hoc way of evaluating
the Galactic prior.  Nevertheless, for completeness they likewise considered 
both cases (with and without the extra factor derived from the
``Rich argument''), and they likewise found that the choice had
only a weak effect on their statistical conclusions.

Here I evaluate analytically (in closed form) the relative likelihood of
discretely degenerate microlens parallax solutions for the case that
the observables are well measured.  I show that the relative 
probability $P_i$ takes the form
\begin{equation}
P_i = K H_i B_i ,
\label{eqn:gamma}
\end{equation}
where $H_i$ may be thought of as the number of Galactic stars with
the physical properties (mass, distance, transverse velocity) of the
$i$-th inferred solution and $B_i = D_{L,i}/\pi_{\e,i}$ is an additional
factor coming from the Jacobian.  The latter should be associated with
the ``Rich argument'', although it differs somewhat from the $\pi_\e^{-2}$
factor that had been originally proposed.  I also discuss how this
exact formula evolves as the assumption of perfect measurement of the 
observables is relaxed.


\section{{Derivation}
\label{sec:derive}}

I assume that the Einstein timescale $t_\e$ and
the angular Einstein radius $\theta_\e$ are precisely measured,
but that the microlens parallax $\bpi_\e$ suffers from a discrete
degeneracy.  While we will be most interested in the case that
each of these local solutions is also precisely determined, it
will also be important to consider that these measurements
have finite, and possibly different, error ellipses (or more generalized
error distributions).  In addition, consideration of these finite
error distributions will allows us to better understand how the
``Rich factor'' behaves in the face of deteriorating errors.

For simplicity, I will initially assume that the source proper motion
$\bmu_s$ and source parallax $\pi_s$ are also known precisely.  While,
this is sometimes true of $\bmu_s$, it is essentially never true of $\pi_s$,
so I will later discuss how the results are affected when these
assumptions are relaxed.

In the usual formulation of the problem, there are then four
observables,
\begin{equation}
(\theta_\e,t_\e,\pi_{\e,N},\pi_{\e,E})
\qquad ({\rm 4\ ``standard''\ observables})
\label{eqn:4standard}
\end{equation}
Hence, for example, when the errors in one or more of these quantities
is poorly constrained (or unconstrained), one carries out a Monte Carlo
simulation of many events drawn from a Galactic model, and one then derives
values and error bars for various physical properties, 
such as the lens mass $M=\theta_\e/\kappa\pi_\e$,
by summing over the simulated events that are consistent with these
measured observables.

A key point of principle is that $(\pi_{\e,N},\pi_{\e,E})$ are not in
fact ``observable''.  Rather what is observed is 
$(\Delta\tau,|u_{0,\sat}|,|u_{0,\oplus}|)$.
Then there are four different combinations
of $(\pi_{\e,N},\pi_{\e,E})$ that are consistent with these observables.
Hence in the Monte Carlo integration imagined in the previous paragraph,
the vast majority of parameter space would contribute essentially nothing
to the integral, while the integrand would be finite in four small regions
where the values of $\bpi_\e$ reproduced the ``true observables'':
$(\Delta\tau,|u_{0,\sat}|,|u_{0,\oplus}|)$.

Nevertheless, for simplicity of exposition, I will treat 
$(\pi_{\e,N},\pi_{\e,E})$ as observables,
keeping in mind that they are a short hand that is applicable 
only locally for the quantities that are actually observed.
I will also substitute 
$\mu = \theta_\e/t_\e$ for $t_\e$ as an observable.  Then the
four observables become
\begin{equation}
(\mu,\theta_\e,\pi_{\e,N},\pi_{\e,E})
\qquad ({\rm 4\ adopted\ observables})
\label{eqn:4adopted}
\end{equation}

In general, if we want to estimate the Bayesian expectation 
of some quantity $Z$, that is a function of observables, we would
evaluate the ratio of integrals
\begin{equation}
\langle Z \rangle = 
{\int d^2\bmu\, d\ln M\, d\ln D_L H(\bmu,D_L,M)\theta_\e \mu E(\bTheta)Z(\bTheta)
\over
\int d^2\bmu\, d\ln M\, d\ln D_L H(\bmu,D_L,M)\theta_\e \mu  E(\bTheta)},
\label{eqn:zbar}
\end{equation}
where 
\begin{equation}
H(\bmu,D_L,M) \equiv f(\bmu)\rho(D_L)D_L^3 \Phi(M)
\label{eqn:hdif}
\end{equation}
is the ``effective number'' of potential lenses with 
Galaxy parameters $(\bmu,D_L,M)$, and 
where $\bTheta\equiv (\mu,\theta_\e,\pi_{\e,N},\pi_{\e,E})$
are the observables, $E(\bTheta)$ is likelihood function of these
parameters derived from the microlensing analysis, 
$(\pi_{\e,N},\pi_{\e,E},\theta_\e)$ are regarded as implicit
functions of the integration variables (keeping in mind that $(\pi_s,\bmu_s)$ 
are considered known), $\Phi(M)= dN/d\ln M$ is the mass function
(normalized to unity),
$\rho(D_L)$ is the number density of stars along the line of sight, 
and $f(\bmu;D_L)$
is the proper motion distribution (normalized to unity).  Then, 
for example, in a numerical
Bayesian analysis, one might simplify $E(\bTheta)$ into a product
of 1-D and/or 2-D Gaussians and then integrate by Monte Carlo.

However, in the present case. we are only interested in the denominator
of Equation~(\ref{eqn:zbar}), i.e., the total probability of a given
solution.  Then, by comparing the probabilities of different discrete
solutions, we can determine their relative likelihood.

First I make a variant of a standard coordinate transformation 
(e.g., \citealt{mb09387})
$(D_L,M)\rightarrow (\pi_\e,\theta_\e)$ by means of the Jacobian,
\begin{equation}
J = \bigg|{\partial (\ln D_L,\ln M)\over\partial (\pi_\e,\theta_\e)}\bigg|
= {2 D_L\over \au}. 
\label{eqn:jacobian}
\end{equation}


Then. the denominator becomes,
\begin{equation}
P = K_1\int d^2\bmu\, d\theta_\e d\pi_\e H(\bmu,D_L,M)D_L \theta_\e \mu 
E(\bTheta),
\label{eqn:p1}
\end{equation}
where $K_1 = 2/\au$.
Next, I implement the assumptions that $\theta_\e$ and $\mu$ are
precisely measured to be $\theta_{\e,0}$ and $\mu_0$, and so write 
$E(\bTheta) = 
\delta(\theta_\e - \theta_{\e,0})
\delta(\mu - \mu_0)
E_\pi(\bpi_\e)$,
where $E_\pi$ is the 2-D likelihood distribution of $\bpi_\e$
in the neighborhood of a given local degenerate solution.
Then Equation~(\ref{eqn:p1}) becomes,
\begin{equation}
P = K_2\int d\phi_\mu\,d\pi_\e H(\bmu,D_L,M) D_L E_\pi(\bpi),
\label{eqn:p2}
\end{equation}
where $\phi_\mu$ is the polar angle associated with $\bmu$ and
$K_2 = K_1\theta_{\e,0}\mu_0^3$.

Then noting that $\bpi_\e$ and $\bmu$ have the same polar angles,
so that $d^2\bpi_\e = d\pi_\e d\phi_\mu \pi_e$, 
Equation(\ref{eqn:p2}) becomes
\begin{equation}
P = K_2\int d^2\bpi_\e H(\bmu, D_L, M) {D_L\over \pi_\e}E_\pi,
\label{eqn:p3}
\end{equation}
or
\begin{equation}
P = K_2\int d^2\bpi_\e \,\eta(\pi_\e) f(\bmu) E_\pi(\bpi_{\e}),
\label{eqn:p3p}
\end{equation}
where
\begin{equation}
\eta(\pi_\e) \equiv \rho(D_L)D_L^4 \Phi(M)/\pi_\e
\label{eqn:etadef}
\end{equation}
is regarded as an implicit function of $\pi_\e$ via
$D_L = \au/(\theta_\e\pi_\e + \pi_s)$ and $M = \theta_\e/\kappa\pi_\e$.

If $E_\pi(\bpi_{\e})$ is taken to be a Gaussian whose effective domain
is small enough that 
$\eta(\pi_\e) = \eta(\pi_{\e,0}) + \eta^\prime(\pi_\e - \pi_{\e,0}) + \ldots$
can be treated as linear in the neighborhood of the solution $\bpi_{\e,0}$,
and $f(\bmu)$ can likewise be treated as linear, then Equation~(\ref{eqn:p3p})
can be directly evaluated:
\begin{equation}
P = K_2[\rho(D_{L,0})D_{L,0}^3 \Phi(M_0)f(\bmu_0)] 
\biggl({D_{L,0}\over \pi_{\e,0}}\biggr)
\label{eqn:p3pp}
\end{equation}
The first term in brackets can be regarded as the naive ``Galactic model
term'', which simply records the local frequency of lenses
with the physical properties inferred from the solutions.  That is,
this is the total number of such lenses $(\rho D_L^3)$ times the fraction 
of such lenses with the inferred mass $M$ and relative proper motion $\bmu$.
The second term in brackets is the suppression
of nearby (small $D_L$, large $\pi_\e$) solutions, i.e., the 
``Rich argument''.  Note that this second factor is stronger (i.e., larger
for larger $D_L$) by $D_L \pi_\e \propto (1 - D_L/D_S)$ (for fixed $\theta_\e$)
than the factor derived using more qualitative arguments by
\citet{21event}.

\section{{Relaxing Assumptions}
\label{sec:relax}}

I had assumed that $(\pi_s,\bmu_s)$ are known
precisely.  The assumption about $\bmu_s$ plays almost no role.
This vector only enters via $f(\bmu)$, 
where $\bmu=\bmu_l - \bmu_s$.  The assumption regarding $f(\bmu)$, 
was only that it was effectively linear in $\bmu$ over the
space of allowed solutions.  For small error bars in 
$\bpi_\e = \bmu (\pi_\e/\mu)$ this is likely to be true even
when $\bmu_s$ is known,  For the case that $\bmu_s$ is not known,
the distribution is more ``smeared out'' and therefore even
more consistent with being linear over small regions.  
By contrast, $\pi_s$ is rarely
if ever known precisely.  More typically, it is estimated with a 10\%
error.  Still, if one makes the evaluation at the average value of
$\pi_s$, then the error in the ratio of $D_L$ terms is quadratic in
the fractional error in $\pi_s$, so no more than a few percent.
This is not likely to enter in a material way into quantitative 
probability arguments.

A more serious issue is the assumption that the joint likelihood distribution
of the two components pf $\bpi_\e$ is Gaussian.  Actually, the argument
given does not really require that it be Gaussian, but only that it
be symmetric in reflections through the best fit $\bpi_{\e,0}$  The
problem is that even this weaker condition is not met for many 
events in the {\it Spitzer} microlensing survey.  In particular,
events for which the {\it Spitzer} data do not cover the peak
(or, more accurately, do not probe an approach to the peak),
the parallax solutions tend to form an arc \citep{gould19}.
In severe cases, the two arcs can even merge to form a large
part of a circle \citep{gould19,kojima2}.  In such cases, it does
not even make sense to talk about separate solutions and so relative
probability of separate solutions.  However, there can be intermediate
cases for which the arcs deviate considerably from an ellipse but still
form two separate solutions.  Depending on the precision required,
one might have to abandon the analytic result given in 
Equation~(\ref{eqn:p3pp}) and carry out a numerical evaluation.
However, one should keep in mind that the default procedure for such numerical 
Bayesian analyses is often to represent the parallax error distribution
as a Gaussian.  If such an approximation is made, the final result
will essentially reproduce the analytic results given here
(unless the error ellipse is so large as to violate the linearity
assumption for $\eta$, defined in Equation~(\ref{eqn:etadef})).  

Another, much more common, deviation from the assumptions of 
Section~\ref{sec:derive} is that $\theta_\e$ is not measured.  This
is relatively rare for planetary and binary events, for which
caustic crossings are usually observed, leading to measurements of $\rho$
and so $\theta_\e = \theta_*/\rho$.  These events are of exceptional
interest and so generally lead to the most detailed investigations.

However, non-planetary (point-lens) events are also very important,
if only because they form the comparison sample for planetary 
events\footnote{Note that this concern does not apply to another class
of non-planetary events: isolated-star mass measurements.  For these,
by definition, $\theta_\e$ is measured.}.
Moreover, some planetary events do not have caustic crossings, and
in these cases $\theta_\e$ is generally not measured.

In this case, the formalism of Section~\ref{sec:derive} can still be
used, but significant caution is required.  The key problem is that in these
cases $D_L$ (for each degenerate parallax solution)
can only be estimated with the aid of a Bayesian analysis.  The
physical basis for such $D_L$ determinations was already recognized by
\citet{han95}: the projected velocities $\tilde v = \au/\pi_\e t_\e$
of disk lenses is roughly proportional to $D_L$, and those of
bulge lenses are very high.  Thus, in most cases, such Bayesian analyses
will yield a relatively well localized distance.  And with this distance
(provided that it is relatively well localized), one can
directly apply Equation~(\ref{eqn:p3pp}).  A ``problem'' with this
approach is that the Bayesian analysis, in addition to yielding a distance
estimate (via Equation~(\ref{eqn:zbar}) with $Z\rightarrow D_L$) will also
yield the probability of this solution (via the denominator of this
same equation).  Moreover, it does so without the simplifying
assumptions that went into the analytic results of Equation~(\ref{eqn:p3pp}).

Nevertheless, even in these cases, Equation~(\ref{eqn:p3pp}) provides
an important sanity check on the results of the numerical integration.
That is, while the logic of the Bayesian integration over a Galactic
model is transparent, its output can be somewhat opaque.  Hence,
it is quite useful to have a simple consistency check on this output.



\acknowledgments

I received support from JPL grant 1500811.

\end{document}